\documentclass[12pt]{revtex4}

\usepackage{amsmath}
\usepackage{amssymb}
\usepackage{psfrag}
\usepackage{graphics}

\def\be{\begin{equation}}
\def\ee{\end{equation}}
\def\bea{\begin{eqnarray}}
\def\eea{\end{eqnarray}}
\def\GB{R_{\textsc{GB}}^2}
\def\a{\alpha}
\def\b{\beta}
\def\g{\gamma}
\def\gk{\gamma_\kappa}
\def\d{\delta}
\def\k{\kappa}
\def\l{\lambda}
\def\r{\rho}
\def\s{\sigma}
\def\t{\tau}

\def\tl{\theta_\lambda}
\def\tm{\theta_\mu}

\def\N{\nabla}
\def\p{\partial}
\def\Om{\Omega}

\begin{document}

\setlength{\unitlength}{1mm}

\title{Unsuccessful cosmology with Modified Gravity Models}

\author{Antonio De Felice\footnote{a.de-felice@sussex.ac.uk}, Mark Hindmarsh\footnote{m.b.hindmarsh@sussex.ac.uk}}

\affiliation{Department of Physics \&\ Astronomy, University of Sussex,
Brighton BN1 9QH, United Kingdom.}

\begin{abstract}
A class of Modified Gravity Models, consisting of inverse powers of linear combination of quadratic curvature invariants, is studied in the full parameter space.  We find that singularity-free cosmological solutions, interpolating between an almost-Friedmann universe at Big Bang Nucleosynthesis and an accelerating universe today, exist only in a restricted parameter space. Furthermore, for all parameters of the models, there is an unstable scalar mode of the gravitational field.
Therefore we conclude that this class of Modified Gravity Models is not viable.
\end{abstract}

\maketitle

\section{Introduction}

One of the most puzzling problems in contemporary physics is the the accelerating expansion of the universe \cite{Riess:1998cb,Perlmutter:1998np,Riess:2004nr, Tonry:2003zg, Spergel:2006hy, Page:2006hz, Hinshaw:2006ia, Jarosik:2006ib, Netterfield:2001yq, Halverson:2001yy}. Standard General Relativity (GR) can accommodate acceleration through a cosmological constant but gives no explanation of its size.

The problem of such a universal constant is difficult to attack with only GR and the Standard Model of Particle Physics (SM). This leads to two possibilities, of which one is to try to see if extensions to the Standard Model motivated by other physics can lead, more or less naturally, to the solution of this problem. Failing that, one can introduce a pure phenomenological model, more or less inspired by some symmetry principles, to try to fit the acceleration data with the fewest free parameters.

Among these models, the most developed one is quintessence \cite{Weiss:1987xa, Wetterich:1987fm, Ratra:1987rm, Peebles:1987ek, Frieman:1995pm, Coble:1996te, Peebles:1998qn, Steinhardt:1999nw, Zlatev:1998tr, Ferreira:1997au, Liddle:1998xm, Copeland:1997et, Ng:2001hs, Bludman:2004az,Linder:2006hh}, that is a minimally coupled scalar field with a properly chosen potential. However a completely different approach has recently emerged, shifting the acceleration behaviour from the matter sector to the gravitational one. In other words, the acceleration, according to these models, is a purely gravitational effect, induced by changing the Einstein-Hilbert action for gravity.

These modifications must become important at late times in order to explain the fact that the universe has started to accelerate only recently, and so much attention has been focused on actions with inverse powers of curvature invariants, mainly squares of the Ricci scalar $R$, the Ricci tensor $R_{\mu\nu}$, and the Riemann tensor $R_{\mu\nu\rho\sigma}$ \cite{Carroll:2003wy, Carroll:2004de, Capozziello:2003tk, Deffayet:2001pu, Freese:2002sq, Arkani-Hamed:2002fu, Dvali:2003rk, Nojiri:2003wx, Arkani-Hamed:2003uy, Abdalla:2004sw, Vollick:2003aw, Mena:2005ta}. These include theories where the Einstein-Hilbert action is supplemented by a function of the Ricci scalar alone, $f(R)$, and also $f(R,P,Q)$ where $P$ and $Q$ are the squares of the Ricci and Riemann tensors respectively. 

The first class of theories, in general, can be mapped to a scalar tensor theory \cite{tourrenc, Wands:1993uu}, and are subject to constraints from both solar system tests of gravity (\cite{Hu:2007nk,Chiba:2006jp,Faraoni:2006hx,Faraoni:2006hx,Allemandi:2005tg}) and cosmology (\cite{Amendola:2006we,Movahed:2007cs,Song:2006ej,Bean:2006up,Amendola:2006kh}). The second kind of theories seem to be inspired by the kind of higher order terms induced by radiative corrections \cite{deBerredo-Peixoto:2004if}. It has been noted that on Minkowski and de Sitter background, these theories would have ghosts unless $P$ and $Q$ appear in the Gauss-Bonnet (GB) combination, that is $f=f(R,Q-4P)$ \cite{Nunez:2004ts, Chiba:2005nz, Navarro:2005gh, Barth:1983hb, Stelle:1977ry, Hindawi:1995an, Boulanger:2000rq, Calcagni:2006ye, Woodard:2006nt, Navarro:2005da}. While allowing the possibility of an accelerating universe today, albeit at the cost of an unexplained small parameter, these theories are afflicted by ghosts, instabilities and superluminal modes in the accelerating background for a large part of the parameter space \cite{DeFelice:2006pg}.

If the late-time behaviour constraints future instabilities, in this paper we address further bounds based on cosmological behaviour in the past history of the universe. We find that the story is no more encouraging.

In half the parameter space, the dynamical equations, expressed in terms of linear combinations of $R^2$ and $Q-4P$, can be shown to possess a separatrix corresponding to a scale factor $a(t) \propto t^p$, with $1/2 < p <1 $, which cannot be crossed. This precludes good cosmological behaviour for these models, since starting from a radiation-dominated universe at nucleosynthesis, with $p\approx1/2$, it becomes impossible for the universe to reach an accelerating phase.

For the remaining part of the parameter space, we have performed a detailed numerical analysis in order to find a background which could mimic GR from Big-Bang Nucleosynthesis (BBN) up to present. Even though it is impractical to integrate the equations of motion for such a long interval of time as the modified Friedmann equation is extremely stiff, we establish that GR-like initial conditions can be imposed and integrated for small time intervals. However, we also find that models with inverse powers of quadratic invariants, for any value of the power-exponent, possess in general classical instabilities during radiation domination. Therefore, we conclude that this class of models cannot be a viable explanation for late-time accelerated expansion.

\section{The models}

We will consider models defined by the following action (see \cite{Carroll:2004de})
\bea 
S&=&M_P^2\int d^4x\,\sqrt{-g}\left[\tfrac12\,R-\frac{\tm\,\mu^{4n+2}}%
  {[a_1\,R^2+a_3\,(Q-4P)]^n}\right]+S_m\\
&=&M_P^2\int d^4x\,\sqrt{-g}\left[\tfrac12\,R-\frac{\tm\,\mu^{4n+2}}{a_3^n}\,\frac1%
  {[b\,R^2+\GB]^n}\right]+S_m \label{e:MGM}
\eea
where $b=a_1/a_3-1$, ($a_3\neq0$), $M_P^2=1/(8\pi G)$, $\tm={\rm sign}(\mu)$, $\GB=R^2-4P+Q$, and 
\be 
Q=R_{\a\b\g\d}\,R^{\a\b\g\d}\ ,\qquad P=R_{\a\b}\,R^{\a\b}\ .  
\ee
For our purposes, it is more convenient to introduce auxiliary scalar fields $\lambda$ and $\phi$ and rewrite the action as
\be
S=M_p^2\int d^4 x\,\sqrt{-g}\left[\left(\tfrac12+\frac{2bn\zeta\,\l}{\phi^{n+1}}\right)R-\frac{bn\zeta\,\l^2}{\phi^{n+1}}-
\frac{(n+1)\,\zeta}{\phi^n}+\frac{n\,\zeta}{\phi^{n+1}}\,\GB\right] ,
\ee
where $\zeta=\tm\,\mu^{4n+2}/a_3^n$. We will also find useful the form
\be
S=M_p^2\int d^4 x\,\sqrt{-g}\left[\bigl(\tfrac12+\chi(\lambda, \phi)\bigr)\,R-U(\lambda,\phi)+\xi(\phi)\,\GB\right] ,\label{action_phi_lambda}
\ee
where
\bea
\chi&=&\frac{2bn\zeta\,\l}{\phi^{n+1}}\, ,\\
U   &=&\frac\zeta{\phi^n}\left[bn\,\frac{\l^2}\phi+n+1\right]=\xi\,\l^2\left[b+\frac{n+1}n\,\frac\phi{\l^2}\right] ,\\
\xi   &=&\frac{n\,\zeta}{\phi^{n+1}}\ .
\eea
 With this choice it is clear that the equations of motion, for any background, become of second order for the two scalar fields $\l,\phi$ and for $g_{\mu\nu}$, namely
\bea
\l&=& R\label{eq:lambda}\, ,\\
\phi&=& b\,R^2+\GB\ ,\label{eq:phi}
\eea
and
\bea
&\left({\textstyle\frac12}+\chi\right)\,R_{\a\b} - \N_\a\N_\b\chi+g_{\a\b}\,\Box\chi-2\,R\,\N_\a\N_\b\xi+ 2\,g_{\a\b}\,R\,\Box\xi & \nonumber \\
& +8R_{(\a\nu}\,\N_{\b)}\N^\nu\xi 
-4\,R_{\a\b}\,\Box\xi-4\,g_{\a\b}\,R^{\r\s}\N_\r \N_\s\xi &\nonumber \\
&- 4\,R_{(\a}{}^{\s\t}{}_{\b)}\,\N_\s\N_\t\xi-\tfrac12\,g_{\a\b}\left[\left({\textstyle\frac12}+\chi\right)\,R-U\right]=4\pi\,T_{\a\b} \ .\label{eq:einstein}&
\eea
Eqs.~(\ref{eq:lambda}) and~(\ref{eq:phi}) are second order differential equations for the metric tensor, where the last one involves both second order derivatives in all the dynamical fields, $\l,\phi,g_{\mu\nu}$. We will suppose that $T_{\a\b}$ is represented by a perfect fluid with two components, radiation and collisionless matter, and that the spacetime interval has the Friedmann-Lema\^{\i}tre-Robertson-Walker (FLRW) form $ds^2 = -dt^2 + a^2(t)(dx^2+dy^2+dz^2)$.

\section{The equations of motion on FLRW}

It is also convenient to perform a change of the time variable, so that
\be
dt=\frac{d\tau}{H_0}\ ,
\ee
and $\tau$ is dimensionless. Denoting dimensionless quantities whose prime derivatives use  $\tau$ with a bar, we have $H= H_0\,\bar H$, $\zeta= H_0^{4n+2}\,\bar\zeta$, and
\bea
&R = H_0^2\,\bar R\, ,\qquad&\l = H_0^2\,\bar \l\, ,\\
&\GB = H_0^4\,{\bar R}_{\rm GB}^2\, ,\qquad&\phi = H_0^4\,\bar \phi\ .
\eea

Having rescaled variables so that they are dimensionless, we may remove the bars on the understanding that $H=a^{-1}da/d\tau$, $R=6(H^2+a^{-1} d^2a/d\tau^2)$ and 
$\GB=24H^2(a^{-1} d^2a/d\tau^2)$.
If we parameterize the FLRW metric in the following way \cite{Mena:2005ta}
\be
ds^2=-\frac{e^{-2u}}{H_0^2\,\b^2}\,dN^2+a_0^2\,e^{2N}\,(dx^2+dy^2+dz^2)\ ,
\ee
where $H=\b\,e^u$ and $N=\ln (a/a_0)$, so that $u(0)=-\ln\b$ (defining the dimensionless parameter $\b$), then one has
\bea
R&=&6\,\b^2\,e^{2u}\,[u'+2]\label{eqr1}\\
\GB&=&24\,\b^4\,e^{4u}\,[u'+1]\ ,\label{eqr2}
\eea
where $u'=du/dN$. From the equations for $\l$ (or $\chi$) and $\phi$ one can find a relation among the dynamical variables, that is \be
\phi=b\,\l^2+4\,\b^2\,e^{2u}\,(\l-6\,\b^2\,e^{2u})\ . \label{uphil} \ee
This means that we need only two independent dynamical variables, which we can freely choose the most convenient ones. Now, we have that
 \be u'=\frac{\dot H}{H^2}\
, \ee 
then $u'<0$, assuming that the universe does not superaccelerate, i.e.\ assuming $\dot
H<0$. It is interesting to note that Eq.~(\ref{uphil}) can be solved for $\beta^2 e^{2u}$, obtaining
 \be
H^2 \equiv \b^2\,e^{2u}=\frac{\l\pm\sqrt{\l^2+6\,(b\,\l^2-\phi)}}{12}\ .  
\label{e:h2}
\ee
The quantity inside the square root is always positive, 
as can be easily verified:
\be
\l^2+6(b\l^2-\phi)=R^2-6\GB=36\,\b^4\,e^{4u}\,u'^2\geq0\ .  \ee
Furthermore, in the past, the universe was decelerating and so $\GB<0$. Therefore we take the positive root in Eq.\ (\ref{e:h2}), from which we learn the dependence of $H$ on the other variables $\l$ and $\phi$.  
Let us consider Eq.~(\ref{eqr1}) for $\l$, which can be written
as
 \be
\l=R=3\b^2\,\frac{de^{2u}}{dN}+\l+\sqrt{\l^2+6\,(b\,\l^2-\phi)}\ .  \ee
By differentiating Eq.\ (\ref{e:h2}) we find 
\be
\tfrac14\,\l'+\tfrac14\,\frac{(1+6b)\l\l'}{\sqrt{\l^2+6(b\l^2-\phi)}}-\tfrac34\,\frac{\phi'}%
{\sqrt{\l^2+6(b\l^2-\phi)}} +\sqrt{\l^2+6(b\l^2-\phi)}=0\ . \label{lambdo} \ee 
From Eq.\ (\ref{eq:einstein}) we can derive the modified  Friedmann equation , which is 
\be 8\,\b^4
e^{4u}\,\xi'+2\,\b^2\,e^{2u}\,(\chi'+\chi)-\tfrac13\,U=\frac\rho{\rho_{\rm
    cr,0}}-\b^2\,e^{2u}\
,\label{Friedo} \ee
 where $\xi'=\xi_\phi\,\phi'$, and $\chi'=2b\,\l\,\xi_\phi\,\phi'+2b\,\xi\,\l'$.

Eqs.~(\ref{lambdo}) and~(\ref{Friedo}) can be combined to give a system of two first-order differential equations from in the variable $\lambda$ and $\phi$:
\bea
A_{11}\,\l'+A_{12}\,\phi'&=&B_1\\
A_{21}\,\l'+A_{22}\,\phi'&=&B_2\ , 
\eea 
where
 \bea
A_{11}&=&1+\frac{(1+6b)\l}{\sqrt{\l^2+6(b\l^2-\phi)}}\\
A_{12}&=&-\frac3{\sqrt{\l^2+6(b\l^2-\phi)}}\\
B_1&=&-4\sqrt{\l^2+6(b\l^2-\phi)}\\
A_{21}&=&\tfrac12\,b\,\xi\,\bigl[\l+\sqrt{\l^2+6\,(b\,\l^2-\phi)}\bigr]\\
A_{22}&=&\tfrac1{12}\,\xi_\phi\,\bigl[\l+\sqrt{\l^2+6\,(b\,\l^2-\phi)}\bigr]^2+\tfrac12\,b\,\l\,\xi_\phi\,
\bigl[\l+\sqrt{\l^2+6\,(b\,\l^2-\phi)}\bigr]\\
B_2&=&\tfrac32\,\bigl[\Om_{m0}\,e^{-3N}+\Om_{r0}\,e^{-4N}\bigr]\notag\\
&&\quad{}+\tfrac12\,U-\tfrac18\,\bigl[\l+\sqrt{\l^2+6\,(b\,\l^2-\phi)}\bigr]\,[1+4b\xi\l]\ .
\eea 
Therefore one has
 \bea
\l'&=&\frac{A_{22}\,B_1-A_{12}\,B_2}{\Delta} \label{e:lambdaEq}\\
\phi'&=&\frac{A_{11}\,B_2-A_{21}\,B_1}{\Delta}\ , \label{e:phiEq}
\eea
 where $\Delta=A_{11}\,A_{22}-A_{12}\,A_{21}$. These are the dynamical equations to solve numerically. The choice of these dynamical variables seem to be a natural one, since the two variables have a clear geometrical and covariant meaning.

\section{Analysis of parameter space}

The action has three free parameters $\zeta,b,n$. From the definition of the variable $\phi$ one can see that
\be
\phi=12 \b^4\,e^{4u}\,[3b\,u'^2+2\,(1+6b)\,(u'+1)]\ ,
\ee
and $u'=-(1+q)$, where $q$ is the deceleration parameter, so
\be
\phi=12\,H^4\,[3b\,q^2-2\,(1+3b)\,q+3b]\ .
\ee
Therefore $\phi$ may vanish for particular values of $q$. It should be noted that if $b<-1/6$, $\phi$ cannot vanish, and remains negative for all values of $q$ and therefore at all times. The dependence on $b$ allows us to divide the parameter space into three regions.

\subsection{$b\geq0$}

The variable $\phi$ vanishes when
\be
q_{1,2}=\frac{3b+1\pm\sqrt{1+6b}}{3b}\ .
\ee
provided that $b>-\tfrac16$, and $b\neq0$. It should be noted that for the particular case $b=0$, there is only one real solution, $q=0$, and 
$\phi = \GB$. For  $b>-1/6$ there are two distinct real solutions. When $\phi$ approaches zero, 
the equation for $\phi$ can be approximated by
\be
\phi'\approx-\frac{18}{n+1}\,\frac b{[1+\sqrt{1+6b}]^2}\,\phi+O(\phi^2)=-\nu\,\phi\ , + O(\phi^2)
\ee
where we have used the fact that $\lambda$ does not vanish at the points where $\phi$ does.
The solution near $\phi=0$ is $\phi=A\,e^{-\nu N}$, from which it is clear that  $\phi$ never crosses zero for finite $N$.
Thus the line $\phi=0$ is a separatrix in the $(\lambda,\phi)$ plane.

Suppose we start at $\phi = 0$. Then a flow is defined by the equation for $\lambda$, which must have constant $q=q_*$. A constant deceleration parameter implies
\be
\dot a=\dot a_*\left(\frac{a_*}a\right)^{\!q_*}\ .
\ee
This has two possible solutions
\bea
&a=a_*\,e^{H_*(t-t_*)}&\quad\textrm{if $q_*=-1$}\, ,\\
&a=a_*\,[1+H_*\,(1+q_*)\,(t-t_*)]^{(1/1+q_*)}&\quad\textrm{if $q_*\neq -1$}\, ,
\eea
that is either an exponential or a power-law $a \propto t^p$ with an exponent defined as $p=p_*=(1+q_*)^{-1}$. 
Thus we learn that the power law expansion parameter $p$ can never cross the value $p_*$. 
Hence for a universe which moves from a radiation-dominated phase $p=1/2$ to an accelerating phase $p>1$ we must not have $p_*$ in the interval $(\frac12,1]$.

Therefore we can rule out all models of the type Eq. (\ref{e:MGM}) with $b\ge0$, since $1/2<1/(1+q_1)\le1$, i.e.\ there is a separatrix between radiation domination and the time of acceleration domination.


\subsection{$b<-\tfrac16$}

For $b < \tfrac16$ we cannot use the argument of the previous section as the $\phi = 0$ separatrix corresponds to power law expansions outside the range $\frac12 < p  < 1$.

Let us instead define a new variable $\kappa = \phi/\l^2$. On dimensional arguments, we might expect that $\k\sim O(1)$. On the other hand let us consider what happens to the equations of motion introducing this ansatz. Since
\be
R=6\,H^2\,(1-q)\ ,
\ee
and $q<1$ during and after radiation domination, then $\tl=1$ and we have
\bea
A_{11}&=&1+\frac{1+6b}{\gk}\\
A_{22}&=&-\frac1{12}\frac{n+1}\k\,\xi\,(1+\gk)\,[1+6b+\gk]\\
A_{12}&=&-\frac3{\l\,\gk}\\
A_{21}&=&\tfrac12\,b\xi\,\l\,(1+\gk)\ ,
\eea
where $\gk=\sqrt{1+6b-6\k}$. The definition of $\kappa$ implies that $\k\leq1/6+b$. Therefore one has for the variable $\Delta$ defined in Eqs.~(\ref{e:lambdaEq}, \ref{e:phiEq})
\be
\Delta=\frac\xi{12}\,\frac{1+\gk}{\gk\,\k}\,\bigl[18b\,\k-(n+1)\,\bigl(\gk+1+6b\bigr)^2\bigr]\ .
\ee
It is important to check where the zeros of $\Delta$ are placed, as the dynamical equations 
(\ref{e:lambdaEq},\ref{e:phiEq}) become singular at $\Delta = 0$.  This means that the derivatives $\lambda'$ and $\phi'$ diverge, and hence a divergent derivative of $R$ and $\GB$.

Denoting the value of $\k$ at which $\Delta$ vanishes as $\k_*$, 
defining $\psi\equiv\phi-\k_*\,\l^2$, 
and expanding $\Delta$ about $\k_*$ (giving $\Delta\approx(\partial_\k\Delta)_{\k_*}\,\psi/\l_*^2+\dots$) 
one has
\be
\psi'\approx\frac{F_2(\phi_*,\l_*,N_*)}\psi\ ,
\label{e:psi}
\ee
assuming that neither $F$ nor $(\partial_\k\Delta)_{\k_*}$ vanish, which has solutions $\psi^2\propto N-N_*$. This says that the solution cannot cross the singularity, because either $N>N_*$ or $N<N_*$. We relegate the exceptional cases in which our assumptions are not valid to Appendix \ref{a:except}.

The equation $\Delta = 0$ has two possible solutions for $\kappa$, which solve the quadratic equation
\be
3\!\left(\frac{3b}{n+1}+1\right)^{\!2}\,\k^2+2(1+6b)\!\left[1+6b-(1+3b)\left(\frac{3b}{n+1}+1\right)\right]\k+3b^2(1+6b)^2=0\ .
\ee
If $b=-\tfrac13\,(n+1)$ then $\k=\k_*=-\tfrac32\,b^2$, otherwise there are, in general, two solutions for $\Delta=0$, namely
\bea
\k_1&=&\frac{b(n+1)|1+6b|}{(1+3b+n)^2}\,\bigl[n-3b+\sqrt{-(1+6b)(2n+1)}\bigr]\\
\k_2&=&\frac{b(n+1)|1+6b|}{(1+3b+n)^2}\,\bigl[n-3b-\sqrt{-(1+6b)(2n+1)}\bigr]\ .
\eea
These solutions are real and negative if $b<-\tfrac16$. 
If $b=-\tfrac16$, then $\k_1=\k_2=0$. It should be noted that 
\be
\k_1\leq\k_2\leq\tfrac16+b\ .
\ee

It is useful to determine the relation between $q$ and $\k$. Using the definitions of $\lambda$ and $\phi$  we find
\be
\k=\frac{3b\,q^2-2(3b+1)\,q+3b}{3(q-1)^2}\ .
\ee
For $b<-\tfrac16$, $\k$ is always negative. According to the evolution of our universe from BBN to the point of change of sign for the acceleration (some time in our past), the variable $q$ smoothly evolves from 1 to 0, that is $1>q\geq0$. In this interval $\k$ monotonically decreases as $q$ increases, and smoothly changes in the interval $-\infty<\k\leq b$. Therefore if $\k_1\leq b$, the universe should have evolved from a radiation dominated universe at BBN to a state with a singularity in a derivative of a curvature invariant. 
Of course this scenario is not consistent with our previous analysis as the universe would not be able to cross in general such singularity. This happens when
\be
\k_1\leq b\qquad{\rm or}\qquad\k_2\leq b\ .
\ee

The inequality $\k_1\leq b$ can be recast in the following form
\be
-(n+1)(1+6b)\sqrt{-(1+6b)(2n+1)}\geq-(1+2n)[9b^2-(n+1)(3b+1)]\ ,
\ee
therefore if $9b^2-(n+1)(3b+1)\geq0$, then this is always verified and there is a past-singularity for all values of $b$ such that
\be
 b\leq\frac{n+1-\sqrt{(n+1)(n+5)}}6\ .
\ee
In the case of
\be
9b^2-(n+1)(3b+1)<0\ ,\qquad{\rm or}\qquad \frac{n+1-\sqrt{(n+1)(n+5)}}6\leq b\leq-\frac16\ ,
\ee
then taking the square we find that the $\k_1\leq b$ is equivalent to saying
\be
(1+3b+n)^2\left[b+\frac{2(n+1)+\sqrt{2(n+1)}}{3(2n+1)}\right]
\left[b+\frac{2(n+1)-\sqrt{2(n+1)}}{3(2n+1)}\right]\leq0\ .
\ee
This is verified by the interval
\be
\frac{n+1-\sqrt{(n+1)(n+5)}}6\leq b\leq\frac{-2(n+1)+\sqrt{2(n+1)}}{3(2n+1)}\ .
\ee
Therefore, in total we have that there is a pathological behaviour for
\be
\label{e:bbar}
b\leq\bar b\equiv\frac{-2(n+1)+\sqrt{2(n+1)}}{3(2n+1)}\ .
\ee
It should be noted that $-\tfrac13<\bar b\leq-\tfrac29$. The inequality $\k_2\leq b$ gives a weaker bound on $b$, with the past-singularity interval $b<-[2(n+1)+\sqrt{2(n+1)}]/[3(2n+1)]<\bar b$.

\subsection{$\bar b<b<0$}

In this region, the differential equation possess no  singular points, and therefore one can try to solve the equations from BBN up to present, to check that there is a viable cosmological solution.
We will assume that above a certain redshift ($N\approx-20$, for BBN) the dynamics of the universe follows that predicted by GR very closely, so that in Eq.\ (\ref{Friedo}) we have 
\be
\frac\rho{\rho_{\rm cr,0}} - \b^2\,e^{2u} \to 0.
\ee 
In this way we know how to impose initial conditions for these cosmological models, and we will require that they be free of ghosts and have a positive square of the propagation speed at the initial ``GR-like'' time.

As $H^2 = \b^2\,e^{2u}$ and $\rho \propto \exp(-4N)$, 
the scale factor $a$ must expand as a power of $\tau$ which is very close to $1/2$: writing $a\propto \tau^{\frac12+\epsilon}$ (recalling that $\tau=H_0\,t$) we find
\be
\lambda = 6\epsilon/\tau^2,\qquad \phi = -3/2\tau^4,
\ee
and hence that $\kappa = \phi/\lambda^2 \ll 0$. Indeed, we checked that for values of $N$ close to BBN, one can choose initial conditions for which we have a GR-like evolution for which $\phi'/\phi\approx-8$ and $\lambda'/\lambda\approx-4$. This can be done as follows
\bea
\l_i&=&6\left[\frac{\ddot a_i}{a_i}+H_i^2\right]\\
\phi_i&=&b\,\l_i^2+24\,H_i^2\,\frac{\ddot a_i}{a_i}\ ,
\eea
where 
\bea
H_i^2&=&\Om_{m0}\,(1+z_i)^3+\Om_{r0}\,(1+z_i)^4+\delta\\
\frac{\ddot a_i}{a_i}&=&\delta-\tfrac12\,\Om_{m0}\,(1+z_i)^3-\Om_{r0}\,(1+z_i)^4\ .
\eea
In the previous 2 equations, the Friedmann equation and the second Einstein equation for the initial conditions have been perturbed by a small parameter $\delta$, equivalent to a tiny cosmological constant. Therefore, at a given redshift, the quantities $\phi_i'/\phi_i$ and $\lambda_i'/\lambda_i$ become functions of $\delta$. Then we numerically found that particular $\delta$ which gives $\lambda_i'/\lambda_i=-4$ with a chosen accuracy. Automatically we also found that $\phi_i'/\phi_i=-8$, satisfying the property that $\kappa=\phi/\lambda^2$ is approximately constant ($\kappa_i\sim-10^{10}$).  In order to find such a $\delta$, and more in general to solve numerically the dynamical equations, one needs multiprecision analysis, as $H^2$ must cancel the usual matter (radiation plus dark matter) up to terms of order of $U\ll H^2$. Along the same lines, the full solution of the differential equations requires time-steps so tiny that the evolution (from BBN or radiation-domination to recombination time) of these backgrounds becomes impractical (at least at these redshifts). Even though we were not able to achieve a full time-evolution for the solution, however we could set up GR initial-conditions for these models and check that background was still GR-like for at least a few thousand time-steps.

On the other hand there are further conditions that should be fulfilled by the propagator of the gravitational field: there should be no ghosts and the square of the propagation speed should be positive \cite{DeFelice:2006pg}. These conditions can be written as follows
\begin{enumerate}
\item Absence of ghost-behaviour for the spin-2 graviton
\be
1 + 4\,b\,\xi\,\l  +8\, \ddot \xi > 0
\ee
\item Absence of classical instabilities
\be
c_2^2>0\,\qquad{\rm where}\qquad c_2^2 = \frac{1 + 4\,b\,\xi\,\l  +8\, \ddot \xi }{1+ 4\,b\,\xi\,\l+8\,H\,\dot \xi} \le 1.
\ee
\item Absence of classical instabilities for the spin-0 modes of the metric
\be
c_0^2>0\,\qquad{\rm where}\qquad c_0^2=1+\frac{32}{3\,Q_1}\,\dot \xi\dot H-\frac{8}{3\,Q_2}(\ddot\xi-\dot \xi\,H),
\label{cond_gen3}
\ee
where
\bea
Q_1=4\,b\,(\dot \xi\,\l+\xi\,\dot \l)+8\,\dot \xi\,H^2\qquad{\rm and}\qquad
Q_2=1+4\,b\,\xi\,\l+8H\dot \xi\ .
\eea
\end{enumerate}


%
We now have sufficient information to study the propagator constraints (see \cite{DeFelice:2006pg} for details) as functions of the initial conditions imposed at redshift $z_i$. 

Noting that  $\dot \lambda \sim H\lambda$, $\dot \xi \sim -H \xi$, and that 
$\lambda \ll H^2$, we find that 
\be
Q_1 \approx 8 \,\dot\xi\,H^2\, ,
\ee
whereas
\be
Q_2\approx 1\,,
\ee
as $\xi\lambda$ and $H^2\xi$ both tend to zero as $\tau \to 0$.
Therefore, since $\ddot\xi\propto H^2\xi$, 
it is easy to see that
\be
c_2^2\approx 1\,\qquad{\rm and}\qquad c_0^2\approx1+\frac{32}3\,\frac{\dot\xi\,\dot H}{8\,\dot\xi\,H^2}
=1-\frac{8}3=-\frac53\, .
\ee
The presence of this classical instability makes the evolution of these models inconsistent with the presence of an early radiation-dominated phase at the era of BBN, and therefore they cannot be accepted as viable models, even if the background could evolve from radiation-domination to matter-domination in a way consistent with today's data.


The result is that the parameter space of the initial conditions at redshift $z_i$, which satisfy all the no-ghosts/instabilities constraints is empty. In particular the scalar modes have an imaginary speed of propagation. This way avoids the formidable problem of solving this system of stiff differential equations. 

\section{Conclusions}

In this paper we have carried out a deeper study of the cosmological behavior of a class of modified gravity actions of the form 
\be
S=M_P^2\int d^4x\,\sqrt{-g}\left[\tfrac12\,R-\frac{\tm\,\mu^{4n+2}}{a_3^n}\,\frac1%
  {[b\,R^2+\GB]^n}\right]+S_m\ .
\ee
The resulting equations were considerably simplified by choosing as independent variables two scalars: the Ricci scalar $R$ and a linear combination of $R^2$ and the Gauss-Bonnet term $\GB$, namely $b\,R^2+\GB$. 
In terms of these two variables, the Einstein equations with matter sources take the form given in 
Eqs.\ (\ref{e:lambdaEq}, \ref{e:phiEq}).

The parameter space divides naturally into three regions, $b<\bar b$, $\bar b<b<0$, and $b>0$, where $\bar b\approx-\tfrac16$ (Eq. \ref{e:bbar}).  Although the equations are very stiff, we can nevertheless distinguish two kinds of pathological behaviour, for a large region of the parameter space. In the region $b>0$ the presence of a separatrix was shown to be inconsistent with a universe whose deceleration parameter $q$ smoothly evolves from 1 at nucleosynthesis to today's negative value. In the region $b<\bar b$, it was proved that for $0<q<1$ the universe should have hit a spacetime singularity, at which $\p_\mu R\, \p^\mu R$ diverges. This also proves the inconsistency of the naive argument which states that $\mu\sim H_0$ implies GR at high redshifts.

The region $\bar b<b<0$ proves to be free of separatrices and singularities. However, a study of the propagators of the scalar and tensor modes in backgrounds which start out close to GR at nucleosysnthesis, shows that in this interval, for positive integer $n$, these models are affected by classical instabilities.

In this way we explored the whole parameter space of the generalized models introduced in \cite{Carroll:2004de}, and we found no possibility for a viable cosmology. In \cite{Mena:2005ta}, the authors explored only the simplest case $n=1$, and using constraints on the recent expansion history from Supernova data, reduced the viable parameter space. In this paper, thanks to the constraints on the scalar and tensor propagators introduced in \cite{DeFelice:2006pg}, we have eliminated the parameter space, leaving no room for these modifications of gravity to be the key component responsible for the acceleration of the universe.

It should be pointed out that the theories we have studied are afflicted by classical instabilities (an incorrect sign in the spatial gradients of perturbations), not ghosts (an incorrect sign in the time derivatives of perturbations). This means that not even introducing a Lorentz-violating cutoff $\Lambda \lesssim 3$ MeV along the lines discussed by Cline et al\ \cite{Cline:2003gs} will help.

\begin{acknowledgments}
We want to thank R.\ Battye, P.\ Mukherjee, K.\ Koyama, A.\ Pilaftsis, and D.\ Wands for useful comments. ADF is supported by PPARC.
\end{acknowledgments}

\appendix

\section{Exceptions}
\label{a:except}
In this Appendix, we briefly discuss the implications of dropping the assumptions made in deriving Eq.\  (\ref{e:psi}).

In principle, one could choose initial conditions at $N_*$ such that the function $F_2$ vanishes at that instant, that is equivalent to solve the following algebraic equation for the initial condition for $\l$
\be
\l_*^{2n+1}-\frac{12\,(\Om_{m0}\,e^{-3N_*}+\Om_{r0}\,e^{-4N_*})}{1+\g_{k_*}}\,\l_*^{2n}+g(\k_*,n,b)=0\ ,
\ee
where $g$ is a function of only $b,n,\k_*$. This algebraic equation has at least one and at most three real solutions. Therefore, for any fixed value of $b,n$, the measure of the subset of the solutions which satisfy these initial conditions is zero in the set of all possible solutions, and we will not study them any further.

For $b<-\tfrac16$, it is impossible that $(\partial_\k^2\Delta)_{\k_*}$ vanishes; but it possible, that for a value of $\k=\tilde\k$, $(\partial_\k\Delta)_{\k_*}=0$. If this happens the previous reasoning discussed in section IV-B regarding the impossibility of crossing the singularity, does not apply, as this time $\psi^3\propto N-N_*$. However, in this last case, $\tilde\k=\k_*$, and this equation gives in turn a relation between $b$ and $n$. In fact one has
\be
b=\tilde b\equiv-\frac2{3+6n}\,[2+2n(n+2)\pm(n+1)\,\sqrt{4n^2+6n+3}]\ .
\ee
Having this constraint on $b$, it is now easier to study the presence of ghosts, classical instabilities, and superluminal modes only in terms of one parameter $n$, as $\zeta$ is fixed by today's data, as already stated in section IV-C. Then, the same result (existence of classical instability, i.e.\ imaginary speed of propagation) applies for $b=\tilde b(n)$, where, also in this case, we chose $N_i=-10$ and $1<n\leq25$).


\begin{thebibliography}{99}


\bibitem{Riess:1998cb}
A.~G.~Riess {\it et al.}  [Supernova Search Team Collaboration],
%
Astron.\ J.\  {\bf 116}, 1009 (1998)
[arXiv:astro-ph/9805201].

\bibitem{Perlmutter:1998np}
S.~Perlmutter {\it et al.}  [Supernova Cosmology Project Collaboration],
Astrophys.\ J.\  {\bf 517}, 565 (1999)
[arXiv:astro-ph/9812133].

\bibitem{Riess:2004nr}
A.~G.~Riess {\it et al.}  [Supernova Search Team Collaboration],
Astrophys.\ J.\  {\bf 607}, 665 (2004)
[arXiv:astro-ph/0402512].

\bibitem{Tonry:2003zg}
J.~L.~Tonry {\it et al.}  [Supernova Search Team Collaboration],
Astrophys.\ J.\  {\bf 594}, 1 (2003)
[arXiv:astro-ph/0305008].

\bibitem{Spergel:2006hy}
  D.~N.~Spergel {\it et al.},
  arXiv:astro-ph/0603449.

\bibitem{Page:2006hz}
  L.~Page {\it et al.},
  arXiv:astro-ph/0603450.

\bibitem{Hinshaw:2006ia}
  G.~Hinshaw {\it et al.},
  arXiv:astro-ph/0603451.

\bibitem{Jarosik:2006ib}
  N.~Jarosik {\it et al.},
  arXiv:astro-ph/0603452.

\bibitem{Netterfield:2001yq}
C.~B.~Netterfield {\it et al.}  [Boomerang Collaboration],
%
Astrophys.\ J.\  {\bf 571}, 604 (2002)
[arXiv:astro-ph/0104460].

\bibitem{Halverson:2001yy}
N.~W.~Halverson {\it et al.},
%
Astrophys.\ J.\  {\bf 568}, 38 (2002)
[arXiv:astro-ph/0104489].


\bibitem{Weiss:1987xa}
N.~Weiss,
Phys.\ Lett.\ B {\bf 197}, 42 (1987).

\bibitem{Wetterich:1987fm}
C.~Wetterich,
Nucl.\ Phys.\ B {\bf 302}, 668 (1988).

\bibitem{Ratra:1987rm}
B.~Ratra and P.~J.~E.~Peebles,
Phys.\ Rev.\ D {\bf 37}, 3406 (1988).

\bibitem{Peebles:1987ek}
P.~J.~E.~Peebles and B.~Ratra,
Astrophys.\ J.\  {\bf 325}, L17 (1988).


\bibitem{Frieman:1995pm}
J.~A.~Frieman, C.~T.~Hill, A.~Stebbins and I.~Waga,
Phys.\ Rev.\ Lett.\  {\bf 75}, 2077 (1995)
[arXiv:astro-ph/9505060].

\bibitem{Coble:1996te}
K.~Coble, S.~Dodelson and J.~A.~Frieman,
Phys.\ Rev.\ D {\bf 55}, 1851 (1997)
[arXiv:astro-ph/9608122].

\bibitem{Peebles:1998qn}
P.~J.~E.~Peebles and A.~Vilenkin,
Phys.\ Rev.\ D {\bf 59}, 063505 (1999)
[arXiv:astro-ph/9810509].

\bibitem{Steinhardt:1999nw}
P.~J.~Steinhardt, L.~M.~Wang and I.~Zlatev,
Phys.\ Rev.\ D {\bf 59}, 123504 (1999)
[arXiv:astro-ph/9812313].

\bibitem{Zlatev:1998tr}
I.~Zlatev, L.~M.~Wang and P.~J.~Steinhardt,
Phys.\ Rev.\ Lett.\  {\bf 82}, 896 (1999)
[arXiv:astro-ph/9807002].

\bibitem{Ferreira:1997au}
P.~G.~Ferreira and M.~Joyce,
Phys.\ Rev.\ Lett.\  {\bf 79}, 4740 (1997)
[arXiv:astro-ph/9707286].

\bibitem{Liddle:1998xm}
A.~R.~Liddle and R.~J.~Scherrer,
%
Phys.\ Rev.\ D {\bf 59}, 023509 (1999)
[arXiv:astro-ph/9809272].

\bibitem{Copeland:1997et}
E.~J.~Copeland, A.~R.~Liddle and D.~Wands,
Phys.\ Rev.\ D {\bf 57}, 4686 (1998)
[arXiv:gr-qc/9711068].

\bibitem{Ng:2001hs}
S.~C.~C.~Ng, N.~J.~Nunes and F.~Rosati,
Phys.\ Rev.\ D {\bf 64}, 083510 (2001)
[arXiv:astro-ph/0107321].

\bibitem{Bludman:2004az}
S.~Bludman,
Phys.\ Rev.\ D {\bf 69}, 122002 (2004)
[arXiv:astro-ph/0403526].

\bibitem{Linder:2006hh}
E.~V.~Linder, Phys.\ Rev.\ D {\bf 74}, 103518 (2006)
[arXiv:astro-ph/0609507].


\bibitem{Carroll:2003wy}
S.~M.~Carroll, V.~Duvvuri, M.~Trodden and M.~S.~Turner,
Phys.\ Rev.\ D {\bf 70}, 043528 (2004)
[arXiv:astro-ph/0306438].

\bibitem{Carroll:2004de}
S.~M.~Carroll, A.~De Felice, V.~Duvvuri, D.~A.~Easson, M.~Trodden and M.~S.~Turner,
Phys.\ Rev.\ D {\bf 71}, 063513 (2005)
[arXiv:astro-ph/0410031].

\bibitem{Capozziello:2003tk}
S.~Capozziello, S.~Carloni and A.~Troisi,
arXiv:astro-ph/0303041.

\bibitem{Deffayet:2001pu}
C.~Deffayet, G.~R.~Dvali and G.~Gabadadze,
Phys.\ Rev.\ D {\bf 65}, 044023 (2002)
[arXiv:astro-ph/0105068].

\bibitem{Freese:2002sq}
K.~Freese and M.~Lewis,
 ``Cardassian Expansion: a Model in which the Universe is Flat, Matter
%
Phys.\ Lett.\ B {\bf 540}, 1 (2002)
[arXiv:astro-ph/0201229].

\bibitem{Arkani-Hamed:2002fu}
N.~Arkani-Hamed, S.~Dimopoulos, G.~Dvali and G.~Gabadadze,
arXiv:hep-th/0209227.

\bibitem{Dvali:2003rk}
G.~Dvali and M.~S.~Turner,
arXiv:astro-ph/0301510.

\bibitem{Nojiri:2003wx}
S.~Nojiri and S.~D.~Odintsov,
%
Mod.\ Phys.\ Lett.\ A {\bf 19}, 627 (2004)
[arXiv:hep-th/0310045].

\bibitem{Arkani-Hamed:2003uy}
N.~Arkani-Hamed, H.~C.~Cheng, M.~A.~Luty and S.~Mukohyama,
JHEP {\bf 0405}, 074 (2004)
[arXiv:hep-th/0312099].

\bibitem{Abdalla:2004sw}
M.~C.~B.~Abdalla, S.~Nojiri and S.~D.~Odintsov,
%
Class.\ Quant.\ Grav.\  {\bf 22}, L35 (2005)
[arXiv:hep-th/0409177].

\bibitem{Vollick:2003aw}
D.~N.~Vollick,
Phys.\ Rev.\ D {\bf 68}, 063510 (2003)
[arXiv:astro-ph/0306630].

\bibitem{Mena:2005ta}
O.~Mena, J.~Santiago and J.~Weller,
arXiv:astro-ph/0510453.



\bibitem{tourrenc}
P.~Teyssandier and P.~Tourrenc,
J.\ Math.\ Phys.\ {\bf 24}, 2793 (1983)

\bibitem{Wands:1993uu}
  D.~Wands,
  Class.\ Quant.\ Grav.\  {\bf 11}, 269 (1994)
  [arXiv:gr-qc/9307034].

\bibitem{Hu:2007nk}
  W.~Hu and I.~Sawicki,
  arXiv:0705.1158 [astro-ph].

\bibitem{Chiba:2006jp}
  T.~Chiba, T.~L.~Smith and A.~L.~Erickcek,
  arXiv:astro-ph/0611867.

\bibitem{Faraoni:2006hx}
  V.~Faraoni,
  Phys.\ Rev.\  D {\bf 74}, 023529 (2006)
  [arXiv:gr-qc/0607016].

\bibitem{Faulkner:2006ub}
  T.~Faulkner, M.~Tegmark, E.~F.~Bunn and Y.~Mao,
  arXiv:astro-ph/0612569.

\bibitem{Allemandi:2005tg}
  G.~Allemandi, M.~Francaviglia, M.~L.~Ruggiero and A.~Tartaglia,
  Gen.\ Rel.\ Grav.\  {\bf 37}, 1891 (2005)
  [arXiv:gr-qc/0506123].

\bibitem{Amendola:2006we}
  L.~Amendola, R.~Gannouji, D.~Polarski and S.~Tsujikawa,
  Phys.\ Rev.\  D {\bf 75}, 083504 (2007)
  [arXiv:gr-qc/0612180].

\bibitem{Movahed:2007cs}
  M.~S.~Movahed, S.~Baghram and S.~Rahvar,
  arXiv:0705.0889 [astro-ph].

\bibitem{Song:2006ej}
  Y.~S.~Song, W.~Hu and I.~Sawicki,
  Phys.\ Rev.\  D {\bf 75}, 044004 (2007)
  [arXiv:astro-ph/0610532].

\bibitem{Bean:2006up}
  R.~Bean, D.~Bernat, L.~Pogosian, A.~Silvestri and M.~Trodden,
  Phys.\ Rev.\  D {\bf 75}, 064020 (2007)
  [arXiv:astro-ph/0611321].

\bibitem{Amendola:2006kh}
  L.~Amendola, D.~Polarski and S.~Tsujikawa,
  Phys.\ Rev.\ Lett.\  {\bf 98}, 131302 (2007)
  [arXiv:astro-ph/0603703].

\bibitem{deBerredo-Peixoto:2004if}
  G.~de Berredo-Peixoto and I.~L.~Shapiro,
  Phys.\ Rev.\ D {\bf 71}, 064005 (2005)
  [arXiv:hep-th/0412249].




\bibitem{Nunez:2004ts}
A.~Nunez and S.~Solganik,
Phys.\ Lett.\ B {\bf 608}, 189 (2005)
[arXiv:hep-th/0411102].

\bibitem{Chiba:2005nz}
T.~Chiba,
JCAP {\bf 0503}, 008 (2005)
[arXiv:gr-qc/0502070].



\bibitem{Navarro:2005gh}
I.~Navarro and K.~Van Acoleyen,
Phys.\ Lett.\ B {\bf 622}, 1 (2005)
[arXiv:gr-qc/0506096].

\bibitem{Barth:1983hb}
N.~H.~Barth and S.~M.~Christensen,
Phys.\ Rev.\ D {\bf 28}, 1876 (1983).

\bibitem{Stelle:1977ry}
K.~S.~Stelle,
Gen.\ Rel.\ Grav.\  {\bf 9}, 353 (1978).

\bibitem{Hindawi:1995an}
A.~Hindawi, B.~A.~Ovrut and D.~Waldram,
Phys.\ Rev.\ D {\bf 53}, 5583 (1996)
[arXiv:hep-th/9509142].

\bibitem{Boulanger:2000rq}
N.~Boulanger, T.~Damour, L.~Gualtieri and M.~Henneaux,
Nucl.\ Phys.\ B {\bf 597}, 127 (2001)
[arXiv:hep-th/0007220].

\bibitem{Calcagni:2006ye}
  G.~Calcagni, B.~de Carlos and A.~De Felice,
  Nucl.\ Phys.\ B {\bf 752}, 404 (2006)
  [arXiv:hep-th/0604201].

\bibitem{Woodard:2006nt}
  R.~P.~Woodard,
  arXiv:astro-ph/0601672.

\bibitem{Navarro:2005da}
I.~Navarro and K.~Van Acoleyen,
arXiv:gr-qc/0511045.

\bibitem{DeFelice:2006pg}
  A.~De Felice, M.~Hindmarsh and M.~Trodden,
   ``Ghosts, instabilities, and superluminal propagation in modified gravity
  JCAP {\bf 0608}, 005 (2006)
  [arXiv:astro-ph/0604154].

\bibitem{Wang:2006ts}
  Y.~Wang and P.~Mukherjee,
  Astrophys.\ J.\  {\bf 650}, 1 (2006)
  [arXiv:astro-ph/0604051].

\bibitem{Cline:2003gs}
  J.~M.~Cline, S.~Jeon and G.~D.~Moore,
  Phys.\ Rev.\ D {\bf 70}, 043543 (2004)
  [arXiv:hep-ph/0311312].

\end{thebibliography}
\end{document}